\documentclass[twocolumn]{aastex62}

\received{2019 January 31}
\accepted{2019 April 15}
\submitjournal{ApJ}

\shorttitle{Long-term Hydrodynamic Simulations on the Planetesimals in MMRs}
\shortauthors{Hsieh \& Jiang}

\begin{document}
\title{Long-term Hydrodynamic Simulations on the Planetesimals Trapped in the First-order Mean Motion Resonances}


\author{He-Feng Hsieh}
\affiliation{Institute of Astronomy, National Tsing Hua University, Hsinchu 30013, Taiwan; hfhsieh@gapp.nthu.edu.tw}

\author{Ing-Guey Jiang}
\affiliation{Institute of Astronomy, National Tsing Hua University, Hsinchu 30013, Taiwan; hfhsieh@gapp.nthu.edu.tw}
\affiliation{Department of Physics, National Tsing Hua University, Hsinchu 30013, Taiwan; jiang@phys.nthu.edu.tw}

\begin{abstract}
The resonant perturbations from planets are able to halt the drag-induced migration,
and capture the inwardly drifting planetesimals into mean motion resonances.
The equilibrium eccentricity of planetesimals in resonances,
and the minimum size of planetesimal that can trigger resonance trapping,
have been analyzed and formulated.
However, the analytical works based on the assumption that the disk is axisymmetric,
which is violated by the asymmetric structures developed by planets.
We perform long-term 2D hydrodynamic simulations to study the dynamics of planetesimals 
in the $j:(j+1)$ first-order exterior resonances,
and reexamine the theoretical expressions.
We find the expression of equilibrium eccentricity underestimates the values for resonances with $j < 5$,
in particular the 1:2 resonance that the underestimation can be $30 - 40\%$.
Within the parameter space we explored,
we find the equilibrium eccentricity and the minimum size are reduced in an asymmetric disk.
The amount of discrepancy in eccentricity depends on the degree of asymmetric structures.
For cases of Earth-sized planets, where the disk is less disturbed,
the planetesimal's eccentricity can reach to the values predicted by our modified expression.
For gaseous planets, however, the eccentricity can be $0.01 - 0.02$ smaller in value.
We find the minimum size is 10 times smaller, and the factor seems to be independent of the planet's mass.
The influences of asymmetric profiles on the eccentricity and the minimum size could
affect the outcome of collisions between resonant and nonresonant planetesimals,
and the amount of planetesimals migrated into the planet's feeding zone.

\end{abstract}


\section{Introduction} \label{sec:intro}

In protoplanetary disks, the dust particles follow the Keplerian orbits, 
but the gas rotates slightly less than the Keplerian velocity due to the support of the radial pressure gradient.
The difference in velocities results in the ``headwind'' experienced by the dust particles
that reduces their angular momenta 
and causes them to spiral toward the central star \citep{1972fpp..conf..211W, 1976PThPh..56.1756A, 1977MNRAS.180...57W}.
When the drifting particles cross the orbit of a large object (e.g., protoplanets or planets), 
a fraction of the particles can be accreted onto the object due to the combined effects of gravitational attraction and gas drag.
This process is known as pebble accretion, and provides a viable way for the rapid growth from protoplanets to planets
\citep{2010A&A...520A..43O,2012A&A...544A..32L}.

On the other hand, the resonant perturbations from the planet are able to counteract the effects of gas drag,
and trap the drifting particles into mean motion resonances (MMRs), before the orbital crossing.
The resonance trapping mechanism was first investigated by \citet{1985Icar...62...16W},
who derived the minimum size that the particles can be trapped in the first-order resonances. 
\citet{1987Icar...70..319P} extended the work to high-order resonances.
For particles smaller than the minimum size, the drag force is too large to be opposed by the resonant perturbation.
The typical values of the minimum size ranges from a few kilometers to subkilometers.

For planetesimals trapped in resonances, their eccentricities are pumped by the resonant perturbations and damped by the gas drag,
and eventually reach the equilibrium eccentricity with moderate values.
The expression of the equilibrium eccentricity was derived by \citet{1985Icar...62...16W},
and confirmed numerically via N-body simulations \citep[e.g.,][]{1993Icar..106..264M}, 
and hydrodynamical simulations for the 1:2 resonance \citep{2007A&A...462..355P, 2012MNRAS.423.1450A}.
The induced eccentricities can cause the overlap of orbits for planetesimals in different resonances,
increase the collision rate between resonant planetesimals or between resonant and nonresonant planetesimals,
and result in the destructive collisions \citep{1985Icar...62...16W}.
The fragments smaller than the minimum size can then pass through the resonances,
and could be accreted by the planet.

The analytical studies on the resonance trapping mechanism assume an axisymmetric gaseous disk; however,
the presence of a planet induces asymmetric structures in the disk,
e.g., single/multiple spiral arms 
\citep[e.g.,][]{1979ApJ...233..857G, 1980ApJ...241..425G, 2016ApJ...832..166L, 2018ApJ...859..118B, 2018arXiv181109628M},
a cavity/gap around the planet 
\citep{1993prpl.conf..749L, 1999ApJ...514..344B},
and the eccentric modes
\citep{2006A&A...447..369K, 2016MNRAS.458.3221T, 2019ApJ...872..184L}.
To study the influences of the asymmetric structures on the resonance trapping mechanism,
hydrodynamics simulations or N-body simulations with a realistic gaseous profile are needed.
On the other hand,
the time the induced eccentricity takes to reach the equilibrium eccentricity can be hundreds to thousands of orbits, 
assuming no damping caused by gas drag \citep{1987Icar...70..319P}.
This indicates that long-term simulations are required.

In this paper, we perform long-term hydrodynamic simulations to study the dynamics of planetesimals 
trapped in the first-order resonances, 
and reexamine the analytical expressions of the equilibrium eccentricity and the minimum size.
The paper is structured as follows.
In Section \ref{sec:theo_model}, we present the drag law adopted in this study, 
and introduce the mechanism of resonance trapping.
The numerical method and the initial condition are described in Section \ref{sec:numer_method}.
The simulation results are presented in Section \ref{sec:results}.
The conclusions and summary are in Section \ref{sec:discussion}.

\section{Theoretical model} \label{sec:theo_model}

\subsection{Drag Law} \label{subset:gas_drag}

The strength of drag force depends on the physical conditions of the dust particle and the surrounding gas.
In cases of spherical particles, the drag force can be expressed in the form
\begin{equation}
\mathbf{F}_{D} = - \frac{1}{2}C_D \pi s^2 \rho_g \Delta V \mathbf{\Delta V},
\end{equation}
where $s$ is the particle radius, $\rho_g$ is the gas mass density,
and $\Delta V$ is the relative velocity between the dust particle and the surrounding gas.
The drag coefficient, $C_D$, is a function of three nondimensional quantities \citep[see, e.g.,][]{1976PThPh..56.1756A}:
\begin{enumerate}
\item The Mach number, $M = \Delta V / c_s$, where $c_s$ is the sound speed.
\item The Knudsen number, $Kn = \lambda / (2 s)$, which is the ratio between the mean free path of gas molecules $\lambda$
and the particle diameter $2s$.
\item The Reynolds number, $Re = 2 s \Delta V / \nu$, where $\nu$ is the gas molecular kinematic viscosity.
\end{enumerate}
In an isothermal disk, the relative velocity is $\Delta V \sim \eta V_K$,
where $V_K$ is the Keplerian velocity,  
\begin{equation}
\eta = - \frac{r}{2 V_K^2 \rho_g} \frac{dp}{dr} = \frac{1}{2} (1 - \gamma) h^2
\end{equation}
is the ratio of the gas pressure gradient to the stellar gravity in the radial direction,
$\gamma$ is the slope of the mass density profile ($\rho_g \propto r^{\gamma}$),
and $h$ is the aspect ratio of the gaseous disk.
On the other hand, the sound speed is in the order of $h$, $c_s = h V_K$, and hence the Mach number $M \sim O(h)$.
In a geometrically thin disk, $h \ll 1$, and we therefore neglect the effects of the Mach number on the drag coefficient.

We adopt the drag law in \citet{1977MNRAS.180...57W}.
For $Kn < 2 / 9$, the drag coefficient is approximated by
\begin{eqnarray}
C_D = \left\{
\begin{array}{ll}
24 Re^{-1}   & \quad  Re \le 1 \\
24 Re^{-0.6} & \quad  1 < Re \le 800 \\
0.44         & \quad  Re > 800. 
\end{array}
\right.
\end{eqnarray}
For $Kn > 2 / 9$, we are in the Epstein regime. 
The drag force is given by
\begin{equation}
\mathbf{F}_{D} = - \frac{4 \pi}{3} s^2 \rho_g \bar{v}_{th} \mathbf{\Delta V},
\end{equation}
where $\bar{v}_{th} = \sqrt{8 / \pi} c_s$  is the mean thermal velocity.
In order to make the drag force be continuous at the boundary ($Kn = 2 / 9$),
the gas molecular viscosity is set to $\nu = \bar{v}_{th} \lambda / 2$.

One useful dimensionless quantity is the Stokes number (or dimensionless stopping time), 
for predicting the behavior of the dust particle influenced by the gas drag.
The Stokes number is defined as
\begin{equation}
St = t_s \Omega_K,
\end{equation}
where $\Omega_K$ is the Keplerian angular velocity, and $t_s$ is the stopping time given by
\begin{equation}
t_s = - \frac{m_d \mathbf{\Delta V}}{\mathbf{F}_D},
\end{equation} 
and $m_d$ is the mass of the dust particle.
The stopping time is the timescale the relative velocity decreases by a factor of $e$.

\subsection{Resonance Trapping} \label{subset:resonance_trapping}
For two bodies in the coplanar orbits, if at least one of the resonant arguments
\begin{eqnarray}
\phi_i & = (p + q) \lambda_o - p \lambda_i - q \varpi_i \\
\phi_o & = (p + q) \lambda_o - p \lambda_i - q \varpi_o  \label{eq:phi_o}
\end{eqnarray}
is in a libration and oscillates about either $0$ or $\pi$, 
we say they are in the $p:(p + q)$ MMR \citep{1999ssd..book.....M}.
The subscripts $i$ and $o$ denote the inner and outer bodies, respectively,
and $\lambda$ is the mean longitude and $\varpi$ is the longitude of pericenter.
The nominal resonance location, defined by
\begin{equation}
\frac{a_i}{a_o} = \left( \frac{p}{p + q} \right)^{2 / 3},
\end{equation}
provides an approximate location for the $p:(p + q)$ exterior resonance.

In this study, we focus on the $j:(j+1)$ first-order exterior resonances.
The equations of motion for the outer body are \citep{1985Icar...62...16W, 1999ssd..book.....M}
\begin{equation} \label{eq:res_a}
\frac{da_o}{dt}\text{(res)} = -(j +1) \mu a_o e_o n_o C(\alpha) \sin\phi 
\end{equation}
\begin{equation} \label{eq:res_e}
\frac{de_o}{dt}\text{(res)} = -\frac{1}{2} \mu n_o C(\alpha) \sin\phi,
\end{equation}
where $a$ is the semi-major axis, $e$ is the eccentricity, 
$\mu = m_i / m_\star$ is the planet-to-star mass ratio, $n$ is the mean motion,
and $\alpha = a_i / a_o$ is the ratio of semi-major axes. 
The function $C$ is derived from the disturbing function:
\begin{equation}
C(\alpha) = (2 j + 1 + \alpha \frac{d }{d\alpha}) b^{(j)}_{1/2}(\alpha) - \frac{1}{\alpha^2} \delta_{1j},
\end{equation}
where $b^{(j)}_{1/2}$ is the Laplace coefficient, and $\delta_{1j}$ is the Kronecker delta function.
For a body in the first-order exterior resonances, 
the stationary point is $\phi = \pi$,
and hence its semi-major axis and eccentricity are pumped by the resonant perturbations.

On the other hand, 
the effects of gas drag on the dynamics of particles in Keplerian orbits were analyzed by \citet{1976PThPh..56.1756A}.
Given the specific force in the form of $f \propto \rho_g \Delta V^2$, 
which corresponds to the Stokes regime with $Re > 800$,
in the planar case their results give
\begin{equation} \label{eq:drag_a}
\frac{da_o}{dt}\text{(drag)} = -2 \frac{a_o}{\tau_0} [e_o \eta + (0.35 - 0.16 \gamma) e_o^3]
\end{equation}
\begin{equation}\label{eq:drag_e}
\frac{de_o}{dt}\text{(drag)} = -0.77 \frac{e_o^2}{\tau_0},
\end{equation}
where $\tau_0 = t_s \eta$.
Here, we assume $e_o > \eta$,
and retain the formula corresponding the case of $e \gg \eta, i$ in \citet{1976PThPh..56.1756A}.

The resonance trapping occurs if the effects of resonant perturbations and gas drag are equal and opposite:
\begin{equation} \label{eq:dadt_all}
\frac{da}{dt}\text{(res)} + \frac{da}{dt}\text{(drag)} = 0 
\end{equation}
\begin{equation} 
\frac{de}{dt}\text{(res)} + \frac{de}{dt}\text{(drag)} = 0.
\end{equation}
The expression of equilibrium eccentricity can then be obtained by solving $de / da \text{(res)} = de / da \text{(drag)}$.
To the leading term in $e$, this yields
\begin{equation} \label{eq:e_eq}
e_{\text{eq}} = 1.14 \left( \frac{\Delta V / V_K}{j + 1} \right)^{1 / 2}.
\end{equation}
Then substituting Eq.\ (\ref{eq:e_eq}) into Eq.\ (\ref{eq:dadt_all}), and letting $\sin\phi = -1$,
the minimum size derived by \citet{1985Icar...62...16W} is
\begin{equation} \label{eq:s_min}
s_{\text{min}} = \frac{\rho_g a_i (\Delta V / V_K)^{1/2}}{3 \rho_d \mu C(\alpha) j^{3 / 2}},
\end{equation}
where the outer body is assumed to be in the Stokes regime with $Re > 800$, and $\rho_d$ is the bulk density.
However, we note there is an error in this expression due to the incorrect
relationship employed by \citet{1985Icar...62...16W}, $n_o / n_i = (j / (j + 1))^{3 / 2}$.
With the correct relationship, $n_o / n_i = j / (j + 1)$,
the factor $j^{3/2}$ in the denominator should be $j^{2 / 3} (j + 1)^{5 / 6}$ instead.
This introduces an overestimation in the minimum size by a factor up to $1.8$ for $j = 1$.
In this study, we ignore this correction,
and refer to the expression in Eq.\ (\ref{eq:s_min}) when mentioning the minimum size.

\section{Numerical method} \label{sec:numer_method}

We perform the simulations using the hydrodynamical code FARGO3D version 1.3 \citep{2016ApJS..223...11B},
with the FARGO algorithm developed by \citet{2000A&AS..141..165M}.
We implement the drag law outlined in Section \ref{subset:gas_drag} to include
the drag force exerted by the gaseous disk onto the planets.
The back reactions onto the gas phase are ignored in this study.

\subsection{Algorithm for Gas Drag} \label{subset:algorithm}

The algorithms for the evolution of planets in FARGO3D are:
\begin{enumerate}
\item Update the velocity due to the disk potential with a time step constrained by the Courant--Friedrichs--Lewy condition, $\Delta t_{\text{CFL}}$.
\item Update the position and velocity, due to the potentials from the star and planets, 
via the fifth-order Runge-Kutta (RK5) integrator several times, with a smaller time step $\Delta t_{\text{RK5}}$. 
The value of $\Delta t_{\text{RK5}}$ is set to $0.2 \times \Delta t_{\text{CFL}}$.
\end{enumerate}
In our approach, the drag force is included in the source terms of velocity,
and updated in real time, in the RK5 integration.
We use the bilinear interpolation to approximate the density and velocity of the gas phase where the planet locates.
In order to accurately simulate the dynamics of planets influenced by the drag force,
the time step used in the RK5 integration must be smaller than the stopping time.
Thus, we calculate the stopping time of all the embedded planets/planetesimals before step 2, 
and choose the minimum value (says $t_{s,\text{min}}$).
The time step for each RK5 integration is then set to
\begin{equation}
\Delta t_{\text{RK5}} = \min(f_{ts} \times t_{s, \text{min}}, 0.2 \times \Delta t_{\text{CFL}}).
\end{equation}
The factor $f_{ts}$ is set to $0.25$ to avoid the velocity difference decreasing
by more than a factor of $e$ due to the drag force in one RK5 integration \citep{2012A&A...546A..18M}.

\begin{figure}[!bt]
\plotone{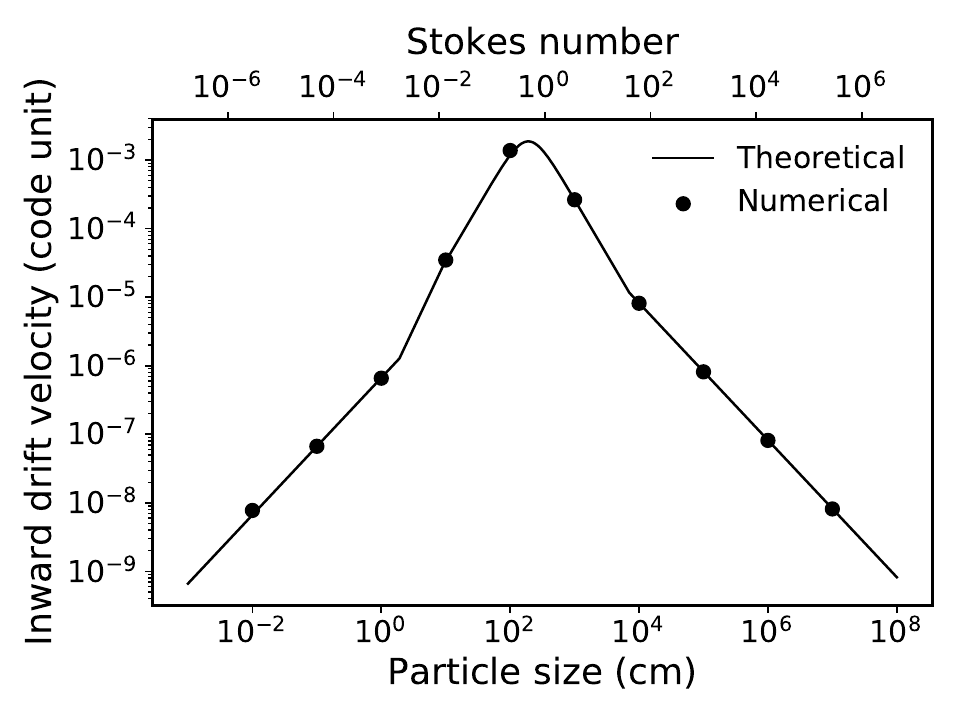}
\caption{Inward drift velocity of dust particles with various sizes,
obtained from the analytical expression in Eq.\ (\ref{eq:drift_velo}) 
and the numerical simulations.\label{fig:code_test}}
\end{figure}

We test our code by comparing the drift velocities of dust particles obtained 
from the analytical expression and the numerical simulations.
In each test run, we embed one particle in an axisymmetric disk.
The particle is initially in a circular Keplerian orbit.
The gaseous disk is assumed to be inviscid to avoid the disk evolution due to the shear force.
We calculate the numerical drift velocity by averaging the simulation results over the last few orbits.
On the other hand, the analytical expression of the drift velocity, $v_{r, d}$, 
is given by \citep{2002ApJ...581.1344T}
\begin{equation}\label{eq:drift_velo}
v_{r,d} = \frac{St^{-1} v_{r,g} - \eta V_K}{St + St^{-1}},
\end{equation}
where the gas radial velocity is set to $v_{r,g} = 0$. 
Figure \ref{fig:code_test} shows the inward drift velocity of dust particles with various sizes,
obtained from Eq.\ (\ref{eq:drift_velo}) and simulations.
Our simulation results agree with the theoretical expectation.

\subsection{Disk Profile} \label{subsec:disk_profile}

We use the 2D fashion of the FARGO3D code.
We assume a locally isothermal, razor-thin disk orbiting a $1~M_\odot$ star, where the aspect ratio is $h = 0.05$.
The disk profile is initially axisymmetric, and extends from $0.2$ to $4.0$ in the code unit.
The unit of length is equal to $1$ au.
The surface density is $1700$ g cm$^{-2}$ at $1$ au, corresponding to the Minimum Mass Solar Nebula model \citep{1981PThPS..70...35H},
and in proportion to $r^{-1}$ inferred from observations \citep{2010ApJ...723.1241A}.
The disk's viscosity is assumed to be uniform, and set to $\nu = 10^{-5}$ in code units, 
which is equivalent to $\alpha = 0.004$ at $1$ au in the alpha-disk prescription \citep{1973A&A....24..337S}.

The disk is resolved by $384$ and $768$ grids in the radial and azimuthal directions, respectively.
We use the damping boundary condition to reduce the unphysical reflecting waves near the boundary.

\subsection{Planet/Planetesimal Profile} \label{subsec:planet_profile}

The planets and planetesimals are all embedded on the midplane of the gaseous disk.
To explore the influences of the asymmetric structures in gaseous disks on the dynamics of planetesimals,
we use three different planet-to-star mass ratios: $\mu = 10^{-3}$, $10^{-4}$, and $10^{-5}$,
which correspond to Jupiter, Neptune, and Earth, respectively.
The planet is fixed in a circular Keplerian orbit, with the orbital radius $a_p = 1$ au.

\citet{1989Icar...82..402D} showed that for a body within the region
\begin{equation}
\frac{|a - a_p|}{a_p} \le \mu^{2/7},
\end{equation}
its behavior becomes chaotic.
For $\mu = 10^{-3}$, the region is $a \in [0.861, 1.139]$ au, 
which is slightly smaller than the nominal resonance location of the 4:5 exterior resonance.
Thus, we focus on the dynamics of planetesimals in $j:(j+1)$ resonances with $j < 5$.

In each simulation, we place one planetesimal in circular orbit for each resonance.
The sizes of planetesimals for simulations with $\mu = 10^{-5}$ are listed in Table \ref{tab:planetesimal_size}.
The sizes are chosen to be slightly larger than the minimum size, based on the values in astronomical unit.
For runs with $\mu = 10^{-4}$ and $10^{-3}$, the sizes are $10$ and $100$ times smaller, respectively.
We also show the corresponding Stokes number in Table \ref{tab:planetesimal_size}.
For a body with a large Stokes number, its dynamic is not strongly perturbed by the gas drag, 
and the drift velocity is small (see Figure \ref{fig:code_test}).
In order to reduce the simulation time required for the drag-induced migration,
the initial orbital radii are slightly larger than the corresponding nominal resonance locations.
We also adjust the locations such that the planetesimals are in near resonances with the embedded planet initially.
In such a compact system, the planetesimals themselves are also in resonances. 
For simplicity, we ignore the interactions between planetesimals.

To calculate the drag force, we assume all the planetesimals are spherical, 
and the bulk density is set to $\rho_d = 1.5$ g cm$^{-3}$.
The mean free path is adopted from \citet{2006MNRAS.373.1619R}:
\begin{equation}
\lambda \sim \frac{4 \times 10^9}{\rho_{g, \text{mid}}} \ \text{cm},
\end{equation}
where $\rho_{g, \text{mid}}$ is the gas mass density on the midplane of the disk.

Note that the force from disk potential is imparted onto the planet's velocity directly in FARGO3D.
For cases of subkilometer-sized planetesimals, the Hill sphere radius is about $10^{-7}$,
which is too small to be used as the smoothing parameter.
We find that the planetesimals are quite unstable in the simulations that employ the Hill radius as the smoothing length.
Therefore, we use the thickness of the gaseous disk as the smoothing length, $\epsilon = 0.4~H$,
where $H$ is the pressure scale height.

\begin{table}[bth!]
\renewcommand{\thetable}{\arabic{table}}
\centering
\caption{The minimum size of planetesimal $s_{\text{min}}$ obtained from Eq.\ (\ref{eq:s_min}),
and the employed planetesimal's size $s_{\text{pl}}$ and the corresponding Stokes number 
for the simulation with an Earth-sized planet, $\mu = 10^{-5}$} \label{tab:planetesimal_size}
\begin{tabular}{cccc}
\tablewidth{0pt}
\hline
\hline
Resonance & $s_{\text{min}}$  & $s_{\text{pl}}$  &  Stokes Number \\
 & ($10^{-8}~\text{au}$/km) & ($10^{-8}~\text{au}$/km) & \\
\hline
\decimals
1:2  &  60.0/89.8 &  100.0/149.6  &  $\sim 10^{6}$ \\
2:3  &  5.4/8.1   &  10.0/15.0    &  $\sim 10^{5}$ \\
3:4  &  2.6/3.9   &  5.0/7.5      &  $\sim 10^{5}$ \\
4:5  &  1.5/2.2   &  5.0/7.5      &  $\sim 10^{5}$ \\
\hline
\end{tabular}
\end{table}

\section{Result} \label{sec:results}

\begin{figure*}[tb!]
\plotone{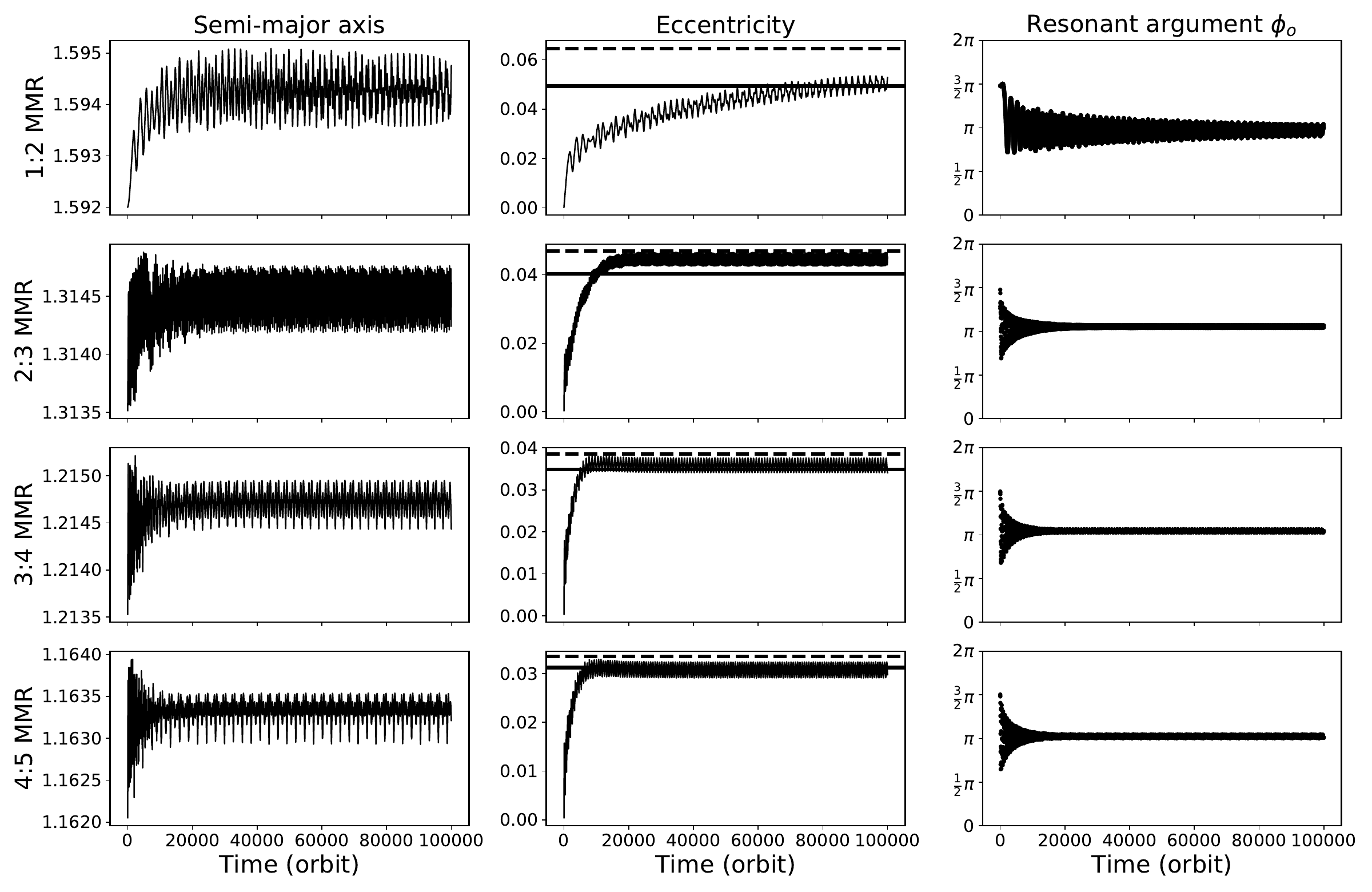}
\caption{Results from hydrodynamic simulations in which the embedded planet is Earth size, $\mu = 10^{-5}$.
From left to right: the time evolution of the semi-major axis, eccentricity, and the resonant argument.
Each row corresponds to a planetesimal in different $j:(j+1)$ exterior resonances.
The horizontal lines in each eccentricity panel are the theoretical values of equilibrium eccentricity:
the solid line is obtained from Eq.\ (\ref{eq:e_eq}),
and the dashed line is from the modified expression in Eq.\ (\ref{eq:e_eq_modified}) that
approximates the numerical solutions of the full equations of motion in \citet{1976PThPh..56.1756A}
(see Section \ref{subsec:ep}).
\label{fig:earth_case}}
\plotone{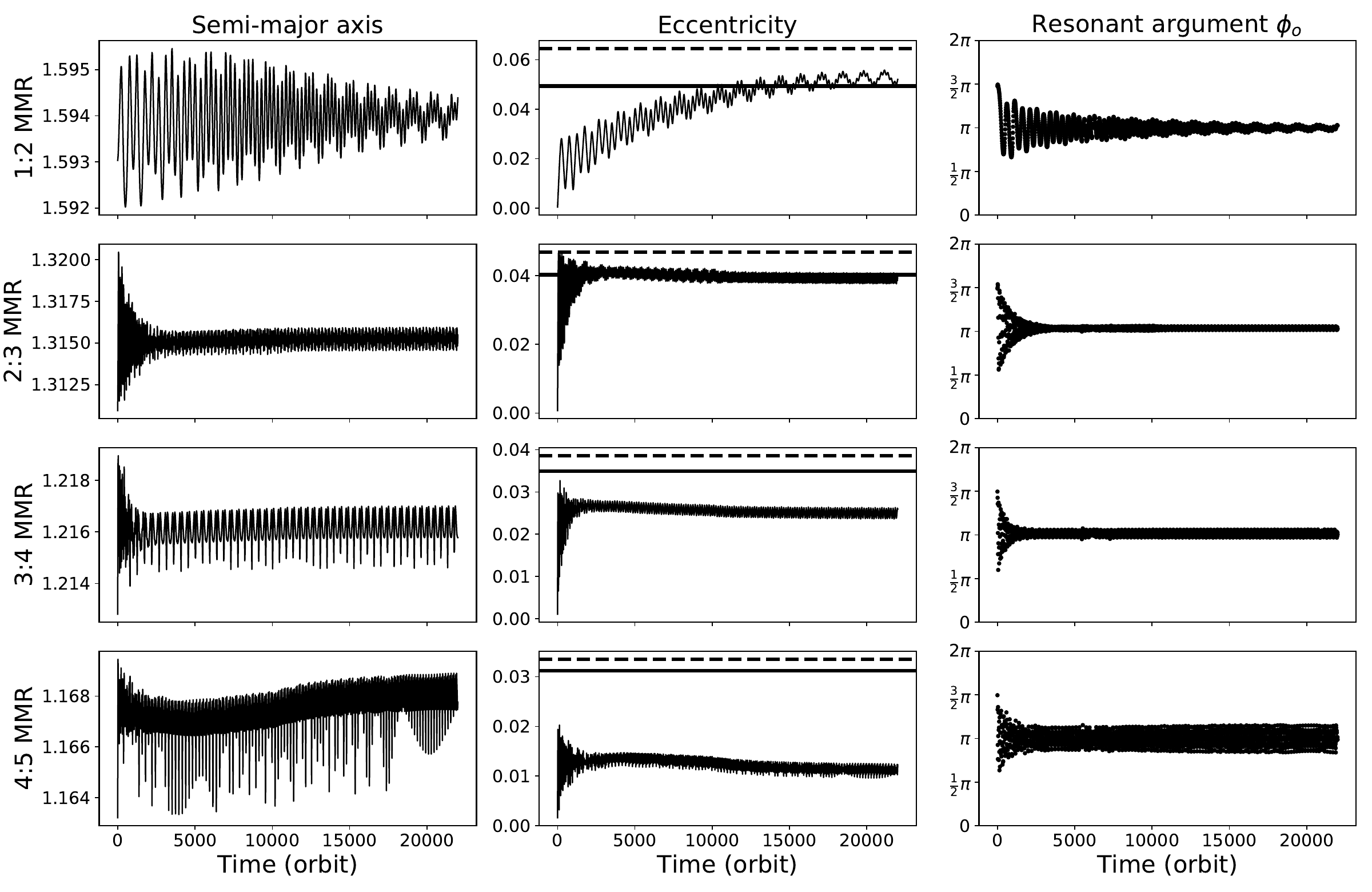}
\caption{Similar to Figure \ref{fig:earth_case}, but for the run with a Neptune-sized planet,
$\mu = 10^{-4}$.\label{fig:neptune_case}}
\end{figure*}

We perform the numerical simulations until the planetesimals are in resonances, 
and their eccentricities reach the quasi-steady state,
except the 1:2 resonance case in the run with $\mu = 10^{-5}$.
Figure \ref{fig:earth_case} shows the results from simulations in which the embedded planet is Earth size, $\mu = 10^{-5}$.
We plot the time evolution of the semi-major axis, eccentricity, and the corresponding resonant argument,
for each planetesimal.
The numerical results for runs with $\mu = 10^{-4}$ and $\mu = 10^{-3}$ are displayed 
in Figures \ref{fig:neptune_case} and \ref{fig:jupiter_case}, respectively.
In the Appendix \ref{appendix_a}, we present the evolution of surface density for the Earth and Neptune cases,
and discuss the effects of viscous evolution on our results.

The simulation time required for the eccentricity reach to the equilibrium state 
strongly depends on the strength of resonant perturbations,
which is a function of the planet-to-star mass ratio $\mu$ and the value of $C(\alpha)$,
ranging from a few hundred orbits for $\mu = 10^{-3}$ 
to dozens of thousands of orbits for $\mu = 10^{-5}$.
For the 1:2 resonance, due to the additional term $\delta_{1j}$ in the function $C$,
the strength of resonant perturbations is much weaker,
and hence the required time is much longer than the other first-order resonances. 

For runs with a Jupiter-sized planet, we find
all the embedded planetesimals, except the one for 1:2 resonance, are accumulated around $r \sim 1.42$.
However, these planetesimals are not in either 2:3 or 3:5 exterior resonances, based on the resonant arguments.
The accumulation of planetesimals is caused by the gap opened by the planet \citep{2007A&A...474.1037F, 2013A&A...552A..66C}.
The pressure bump near the outer edge of the gap provides 
the positive pressure gradient to stop the planetesimals from drifting inward,
which is known as the dust filtration mechanism \citep{2006MNRAS.373.1619R}.

\begin{figure*}[t!]
\plotone{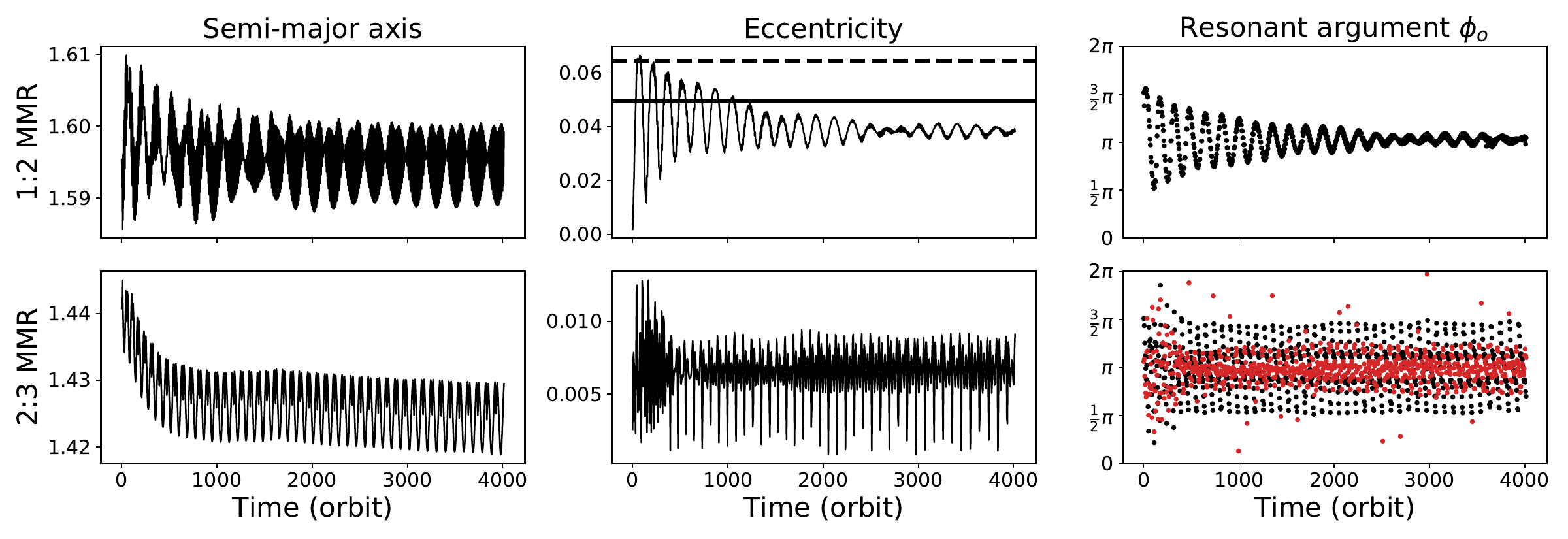}
\caption{Similar to Figure \ref{fig:earth_case}, but for the run with a Jupiter-sized planet, $\mu = 10^{-3}$.
Due to the pressure bump near the outer edge of the gap, 
all planetesimals, except the one for 1:2 resonance, are halted at $r \sim 1.42$.
In the bottom panels, we show the orbital elements of the planetesimal for 2:3 resonance as an example,
in which the resonant arguments of 2:3 resonance and 3:5 resonance are shown in black and red dots, respectively.
\label{fig:jupiter_case}}
\end{figure*}

\subsection{Semi-major Axis} \label{subsec:ap}

We note that the semi-major axes of planetesimals in MMRs are slightly larger than 
the corresponding nominal resonance locations, in all the simulation results.
The outward shifting in the resonance location is caused by the precession of the planetesimal,
which is induced by the external forces, e.g., 
the resonant perturbations from the planet, gas drag,
and the gravitational forces from an axisymmetric gaseous disk \citep{2015ApJ...805..100T}.
At the stationary point where $\phi_o$ remains unchanged,
the exact resonance location can be derived from Eq.\ (\ref{eq:phi_o}) by setting $d\phi_o / dt = 0$,
which yields
\begin{equation}
\frac{a_o}{a_i} = \left( \frac{j + 1 - \dot\varpi_o / n_o}{j} \right)^{2 / 3}.
\end{equation}
When the precession rate is negative (regression, $\dot\varpi_o < 0$), 
the exact resonance location will be larger than the nominal location naturally.

The amount of deviation in the semi-major axis is a function of $\dot\varpi_o / n_o$,
which depends on the strength of the external forces, 
and the radial gradient of the forces 
(see Eq.\ (\ref{eq:eq_varpi_n}) for cases of the gravitational force from the disk).
To gain insight into the main contributor of the precession rate,
we calculate the amount of shifting, $\Delta$, defined as 
\begin{eqnarray}
\Delta &=& \frac{(a_2 / a_1)_{\text{simulation}}}{(a_2 / a_1)_{\text{nominal}}} \\
       &\sim& 1 - \frac{2}{3}\frac{\dot\varpi_o / n_o}{j + 1}.
\end{eqnarray}
In Table \ref{tab:shift_in_a}, we list the values of $\Delta$ obtained from the simulation results.
First, for the 1:2 resonance, we note all the deviations in semi-major axis are about $0.4 \%$,
and seem to be independent of the planet's mass and the size of planetesimal.
Moreover, for the run with $\mu = 10^{-5}$, the deviation decreases with $j$.
These results infer the drag force may not be the main contributor
because of its strong dependence on the particle size and the gas density.
For the resonances with $j > 1$, the deviation increases with the planet's mass, 
and, interestingly, increases with $j$ in cases of $\mu = 10^{-4}$, 
which is contrary to the cases of $\mu = 10^{-5}$.
This may be related to the asymmetric structures induced by the planet, e.g., density wakes.
The planetesimals in MMRs with higher $j$ are more closer to the planet, 
and hence the effects of asymmetric structures become more prominent.

Now we discuss this question quantitatively.
The values of $\dot\varpi_o / n_o$ required for the deviations listed in Table \ref{tab:shift_in_a}
are about $-10^{-2}$.
For the gravitational force from an axisymmetric disk, 
we adopt the Eq.\ (B10) in \citet{2015ApJ...805..100T} to estimate its contribution:
\begin{equation}  \label{eq:eq_varpi_n}
\frac{\dot\varpi}{n} = -\left[ \frac{F_D}{F_K} + \frac{1}{2} r \frac{d F_D / d r}{F_K} \right],
\end{equation}
where $F_K$ is the Keplerian force from the star, and $F_D$ is the radial force from the disk.
The value of $F_D$ is calculated using two methods outlined in \citet{2016A&A...589A.133F}:
the Binney and Tremain approach and the Ward's approach.
The results from these two methods are in the same order, with values of $-10^{-3}$.
Thus, we can neglect the contribution from the axisymmetric disk.
On the other hand, the contribution from the resonant perturbations are \citep{1999ssd..book.....M}
\begin{equation}
\frac{\dot\varpi_o}{n_o} = 2 \mu f_{s, 1}(\alpha) + \frac{\mu}{2e_o}  C(\alpha) \cos\phi_o,
\end{equation}
where
\begin{equation}
f_{s, 1}(\alpha) = \frac{1}{8} \left[2 \alpha \frac{d }{d\alpha} + \alpha^2 \frac{d^2 }{d\alpha^2}  \right] b^{(0)}_{1/2}(\alpha).
\end{equation}
Note that $f_{s,1}$ and $C$ are both positive and increase monotonically with $\alpha$.
For exterior resonances, the resonant argument oscillates around $\pi$, 
and hence the second term is negative such that $\dot\varpi_o / n_o$ could be negative.
To calculate $\dot\varpi_o / n_o$, we simply assume $\cos\phi = -1$.
The values of $e_o$ are set to $0.001$ for $\mu = 10^{-5}$ and $0.01$ for $\mu = 10^{-3}$,
based on the evolution equation of eccentricity that 
the eccentricity increases more slowly for smaller $\mu$ and is expected to be lower. 
Under these assumptions, the values of $\dot\varpi_o / n_o$ contributed from the resonant perturbations 
can reach $-10^{-2}$, or even larger, $-10^{-1}$.
Therefore, we think that in our simulations the deviation in the semi-major axis
is mainly caused by the resonant perturbations,
and possibly the gravitational forces from the asymmetric structures in the gaseous disk.

On the late stage, the semi-major axis increases slowly with time,
accompanied by the decrease in the eccentricity 
(see the cases of MMRs with $j \ge 2$ in Figures \ref{fig:earth_case} and \ref{fig:neptune_case}).
The slow divergent evolution is known as the resonant repulsion mechanism 
\citep{2010MNRAS.405..573P, 2011CeMDA.111...83P, 2012ApJ...756L..11L},
caused by the tidal circularization due to external dissipative forces, 
e.g., the tidal friction from the central star \citep{2013AJ....145....1B},
and the wake--planet interaction \citep{2013ApJ...778....7B}.

\begin{table}[!bth!]
\renewcommand{\thetable}{\arabic{table}}
\centering
\caption{The ratio of the numerical location and the nominal resonance location, $\Delta$, 
in different first-order resonances and planet-to-star mass ratios.} \label{tab:shift_in_a}
\begin{tabular}{c|ccc}
\tablewidth{0pt}
\hline
\hline
$\Delta$ & $\mu = 10^{-5}$ & $\mu = 10^{-4}$ & $\mu = 10^{-3}$ \\
\hline
\decimals
1:2  &  $1.0044$  &  $1.0042$  &  $1.0048$  \\
2:3  &  $1.0031$  &  $1.0037$  & - \\
3:4  &  $1.0027$  &  $1.0039$  & - \\  
4:5  &  $1.0025$  &  $1.0059$  & - \\
\hline
\end{tabular}
\end{table}

\subsection{Equilibrium Eccentricity} \label{subsec:ep}

In Figures \ref{fig:earth_case}--\ref{fig:jupiter_case}, 
we plot the equilibrium eccentricity predicted by \citet{1985Icar...62...16W} 
in each eccentricity panel for comparison (the solid line).
For the 1:2 resonance, 
our results roughly agree with the theoretical prediction.
However, the final eccentricity in the Jupiter cases is about $0.04$,
which is smaller than the theoretical value $0.05$.
The theoretical value of the 1:2 resonance has been examined 
and reported to be consistent with the results from hydrodynamical simulations 
in \citet{2007A&A...462..355P} and \citet{2012MNRAS.423.1450A}.
However, we note that the simulation time in their simulations are about a few hundred orbits,
where the planetesimal's eccentricity in our simulations still oscillates around $0.03 - 0.06$, 
or is not be damped to the equilibrium state
(see Figures \ref{fig:jupiter_case} and \ref{fig:jupiter_smin}).
Thus, the values they reported may not be the equilibrium eccentricity,
but the averaged eccentricity of the evolving pebbles near the 1:2 resonance.
For the resonances with $j > 1$, in the Earth case, 
our results roughly agree with the analytical prediction, 
albeit the value is slightly higher in the case of the 2:3 resonance.
In the Neptune case, however, we find the eccentricity significantly departs from the theoretical values
for planetesimals in the 3:4 and 4:5 resonances (see Figure \ref{fig:neptune_case}).

Recall the expression of equilibrium eccentricity 
is derived with the assumptions that the drag force is the Stokes drag with $Re > 800$,
and the gaseous disk is axisymmetric.
In cases of $1 < Re < 800$, the drag force is stronger, and hence
the equilibrium eccentricity should be smaller than the value predicted by Eq.\ (\ref{eq:e_eq}).
To examine whether the discrepancy is caused by the drag law,
we perform one additional simulation for the Neptune case (not shown here),
in which only the Stokes drag with $Re > 800$ is employed.
However, the final eccentricities are similar to the results shown in Figure \ref{fig:neptune_case}.
We also follow \citet{1976PThPh..56.1756A} to derive the equations of motion 
for the cases of Stokes drag with $1 < Re < 800$,
where the specific force is in the form of $f \propto \rho_g^{0.4} \Delta V^{1.4}$.
The expression of equilibrium eccentricity is similar to Eq.\ (\ref{eq:e_eq}),
but with a smaller coefficient $1.06$ instead.\footnote{
The derivations are skipped in this paper.}
However, this correction is not enough to explain the final eccentricities in the Neptune case.
We therefore rule out the possibility of drag law.

\begin{figure*}[tb!]
\includegraphics[width=\textwidth]{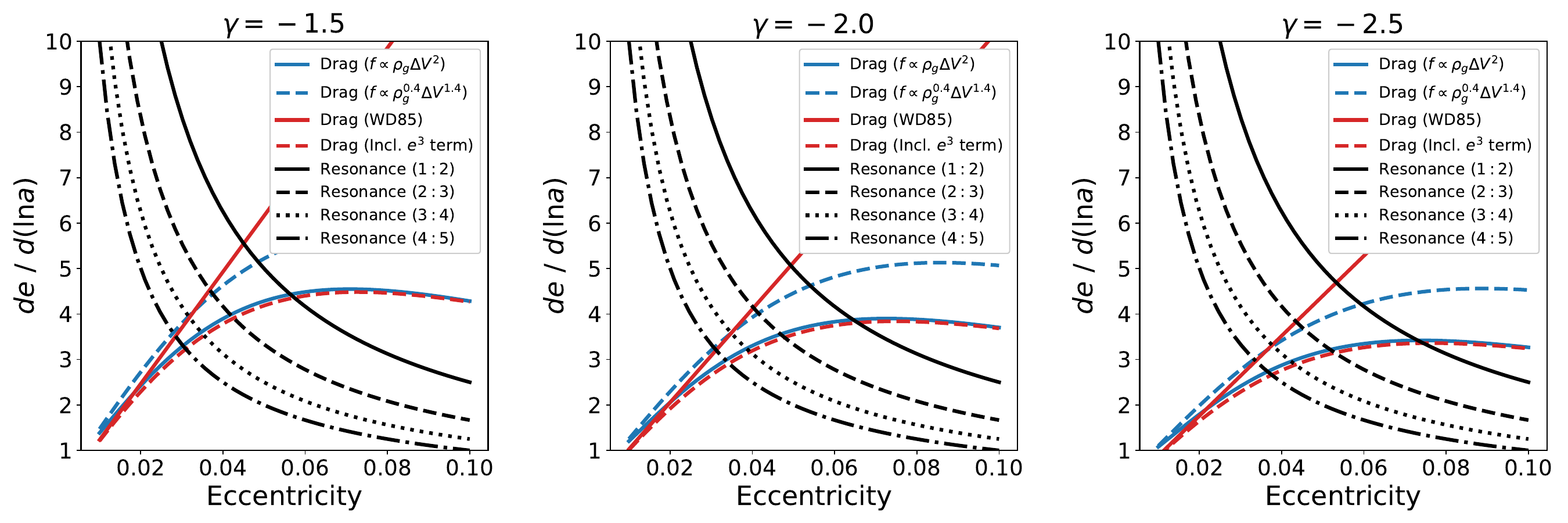}
\caption{Ratio between the equations of motion, $de / d\ln a$, 
in function of the eccentricity, given the slope of mass density $\gamma$ ($\rho_g \propto r^{\gamma}$).
We show the results with $\gamma$ that are frequently employed: 
in values of $-1.5$ (left), $-2.0$ (middle), and $-2.5$ (right).
The relation between $\gamma$ and the slope of surface density is $\Sigma \propto r^{1 + \gamma}$.
The black lines are the values for resonant perturbations in different MMRs,
while the blue lines are the numerical solutions in cases of different form of gas drag,
and the red lines are the approximation used in \citet{1985Icar...62...16W} and this work.
The intersections of these lines are therefore the theoretical values of the equilibrium eccentricity.
\label{fig:numerical_Adachi}}
\end{figure*}

In order to investigate whether the discrepancy in eccentricity is caused by
the asymmetric structures in the gaseous disk, 
we numerically solve the equations of motion in \citet{1976PThPh..56.1756A}
to include all the contributions from the high-order terms in $e$ and $\eta$.
We then calculate the values of $de / d\ln a$ from the equations of motion
for both resonant perturbations and gas drag.
The results are illustrated in Figure \ref{fig:numerical_Adachi}, 
for the resonant perturbations in different MMRs and the gas drag in different forms.
The theoretical values of the equilibrium eccentricity are therefore the intersections
where $de / d\ln a \text{(res)} = de / d\ln a \text{(drag)}$.
From Figure \ref{fig:numerical_Adachi}, 
we note the numerical solutions of equations in \citet{1976PThPh..56.1756A} 
significantly depart from the approximated form used in \citet{1985Icar...62...16W} 
on the high-eccentricity end,
inferring larger equilibrium eccentricities for MMRs with $j < 5$,
in particular the case of $j = 1$.
The underestimation in equilibrium eccentricity has been shown in \citet{1993Icar..106..264M}, 
in which his results from N-body simulations give higher eccentricities in cases of $j \le 5$ (see Figure 4 therein).
We find the departure is caused by the $e^3$ term in $da/dt$ (see Eq.\ (\ref{eq:drag_a})), 
which becomes significant when $e^2 \sim \eta$, 
but ignored in \citet{1985Icar...62...16W}.
Including the contribution from the $e^3$ term,
the equilibrium eccentricity is given by 
\begin{equation} \label{eq:e_eq_modified}
e_{\text{eq}} = \left[ \frac{\Delta V / V_K}{0.77(j + 1) - (0.35 - 0.16 \gamma)} \right]^{1 / 2},
\end{equation}
which approximates the numerical solution with an overestimation up to $3\%$ for $j \le 5$.

The equilibrium eccentricity predicted by Eq.\ (\ref{eq:e_eq_modified}) 
are presented by the dashed lines in Figures \ref{fig:earth_case} to \ref{fig:jupiter_case} for comparison.
With the values from the modified formula,
we find it is easier to explain the final eccentricity in our simulation results,
albeit not exactly match.
Under the influences of asymmetric structures in the gaseous disk,
the equilibrium eccentricity will be smaller than the theoretical values.
The amount of discrepancy strongly depends on the degree of asymmetric structures,
and hence on the planet's mass and $j$.
For cases of Earth-sized planets and protoplanets, 
where the disk is not highly disturbed,
the planetesimal's eccentricity can almost reach to the theoretical values given by the modified expression,
with a negligible discrepancy (see Figure \ref{fig:earth_case}).
For gaseous planets, 
the discrepancy increases quickly with $j$ and the planet's mass,
and becomes significant in cases of $j \ge 3$ for Neptune-sized planets and $j = 1$ for Jupiter-sized planets
(see Figures \ref{fig:neptune_case} and \ref{fig:jupiter_case}).
Lastly, the discrepancy in eccentricity, caused by the asymmetric structures,
could lead to the agreement between the simulation results
and the analytical expression in \citet{1985Icar...62...16W},
as the lucky cases show in our simulations,
e.g., the 3:4 and 4:5 resonances in the Earth case, 
and the 1:2 and 2:3 resonances in the Neptune case.

Note the final eccentricities shown in our results are in the quasi-steady state.
Even if the equilibrium eccentricity is reached, 
the planetesimal's eccentricity will decrease slowly with time,
due to the resonant repulsion mechanism mentioned in Section \ref{subsec:ap}.

\begin{figure*}[tbh!]
\includegraphics[width=\textwidth]{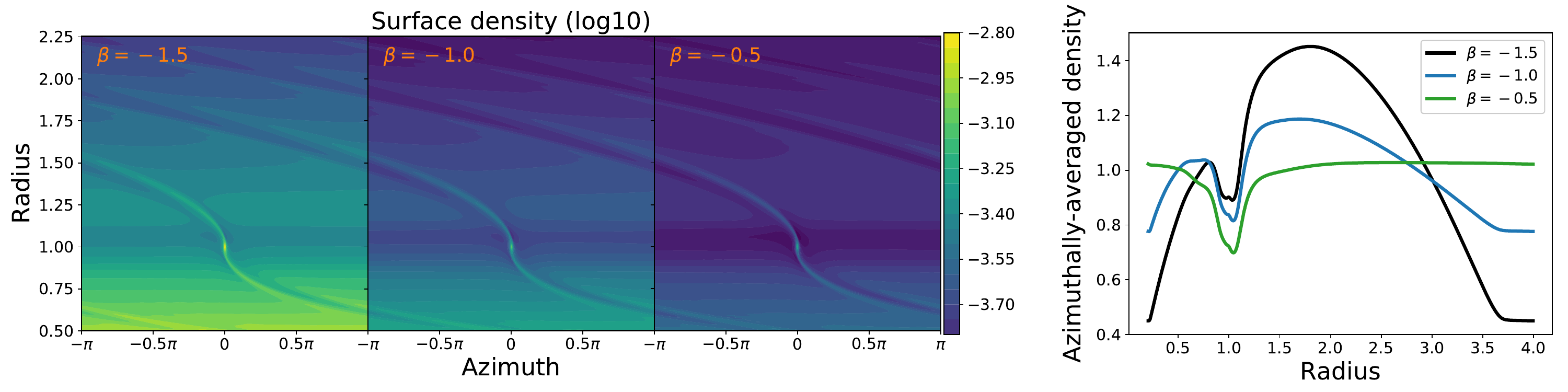}
\caption{Surface density in logarithmic scale (left) and the azimuthally averaged
surface density (right) for runs with different slopes of surface density,
$\beta$ ($\Sigma \propto r^\beta$).
In the right panel, the surface density is rescaled by the total disk mass 
and normalized by the initial value as follows: 
$(\Sigma / M) / (\Sigma_{\text{init}} / M_{\text{init}})$.
With a flatter density profile initially (larger $\beta$), 
the gap opened by the Neptune-sized planet is deeper and wider,
and the asymmetric structures are more prominent.
\label{fig:surface_slope}}
\end{figure*}

\begin{figure*}[!tb]
\plotone{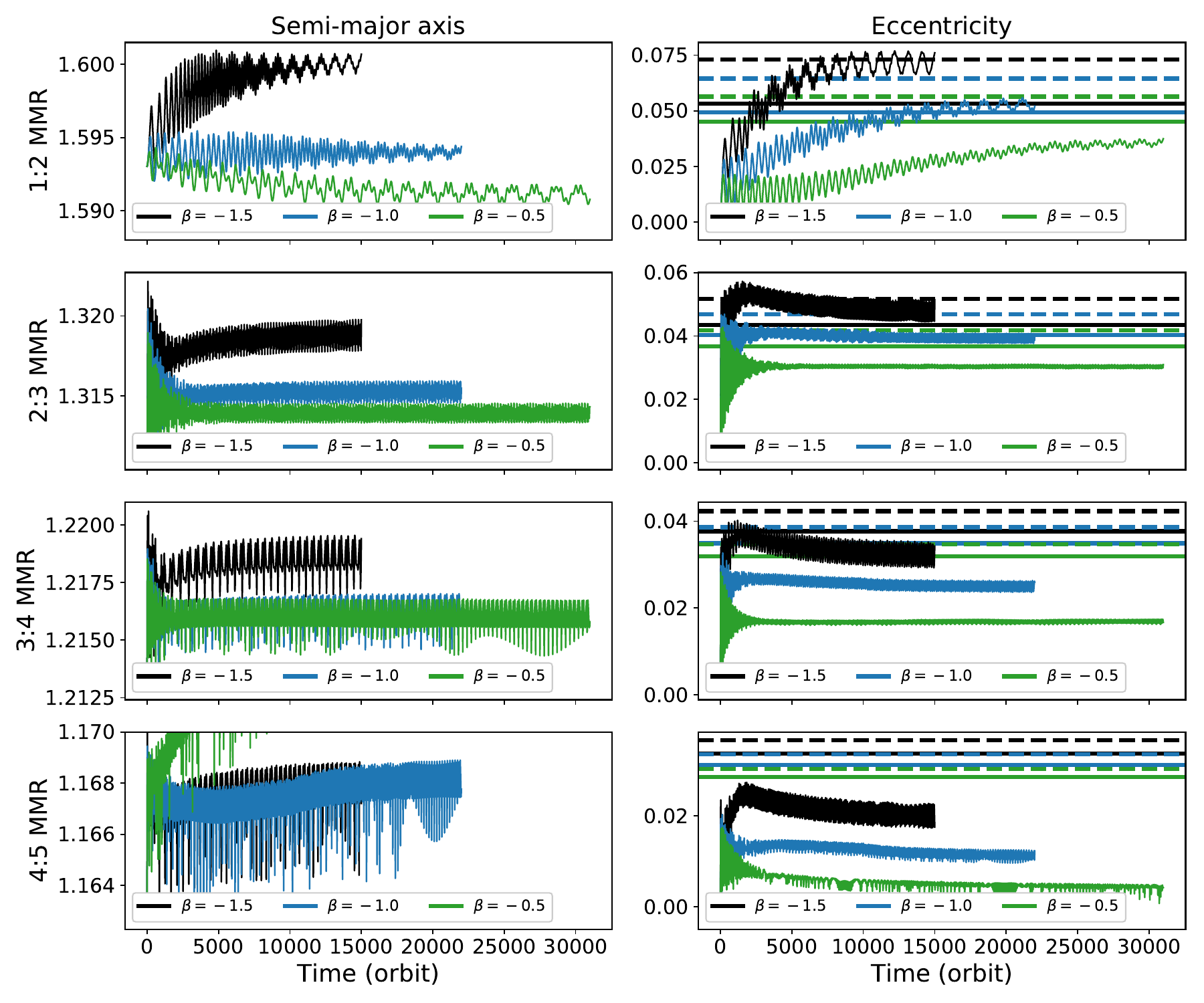}
\caption{Similar to Figure \ref{fig:neptune_case} for runs with a Neptune-sized planet,
but the slope of surface density is varied.
The results from simulations with $\beta = -1.5$ are presented by the black lines,
while $\beta = -1.0$ is in blue, 
and $\beta = -0.5$ is in green.
The results shown in Figure \ref{fig:neptune_case} are included for comparison (the blue lines).
The gap opened by the planet is deeper and wider in the run with $\beta = -0.5$,
in which the the outer edge is about $1.2$. 
As a result, the planetesimal for the 4:5 MMR (the green line) migrates outwardly 
due to the dust filtration mechanism.
\label{fig:orbelem_slope}}
\end{figure*}

\subsection{Dependence on the Slope of Surface Density} \label{subsec:dens_slope}

We mentioned that the analytical expression in \citet{1985Icar...62...16W} underestimates the equilibrium eccentricity, 
and the underestimated values could be consistent with the simulation results occasionally, 
for cases of MMRs with $j < 5$.
To verify our statements,
we perform additional simulations with different initial conditions of the gaseous disk.
We note the analytical expression of equilibrium eccentricity 
depends on the relative velocity $\Delta V$,
which is a function of the density slope and the aspect ratio of the gaseous disk.
To keep the assumption that the disk is razor thin, 
we thus perform additional simulations in which the slope of surface density, $\beta$,
is changed to $-1.5$ and $-0.5$ ($\Sigma \propto r^\beta$), 
with a Neptune-sized embedded planet.

Figure \ref{fig:surface_slope} shows the final profiles of surface density in logarithmic scale.
For a gaseous disk with a flatter density profile initially (larger $\beta$),
the asymmetric structures induced by the planet are more prominent, e.g.,
the larger radial density contrast around density wakes,
and the deeper and wider gap.
In the run with $\beta = -0.5$, the outer edge of the gap reaches to $1.2$,
and hence prohibits the resonance trapping for $j \ge 4$.

The orbital elements of planetesimals are illustrated in Figure \ref{fig:orbelem_slope}.
All the planetesimals are in resonances,
except the one for 4:5 resonance in the run with $\beta = -0.5$ (the green line).  
For the run with $\beta = -1.5$, 
we find the eccentricity of the planetesimal in 1:2 resonance can reach $0.07$,
which is consistent with the modified expression Eq.\ (\ref{eq:e_eq_modified}),
but about $40\%$ higher than the theoretical value predicted by Eq.\ (\ref{eq:e_eq}).
On the other hand, 
the equilibrium eccentricities in cases of the 1:2 and 2:3 MMRs do not agree with 
Eq.\ (\ref{eq:e_eq}) in runs with different $\beta$.
These confirm that the analytical expression in \citet{1985Icar...62...16W} 
underestimates the equilibrium eccentricity,
and could be misleading,
especially for cases of $j \le 2$.

Now we focus on the discrepancy in eccentricity between the simulation results
and the values predicted by the modified expression (the dashed lines).
From Figure \ref{fig:orbelem_slope},
the amount of discrepancy increases with $j$ for all cases of $\beta$,
as also seen in the simulations with different planets' masses.
We find the amount also depends on the initial density slope.
The difference is much larger in the run with a larger $\beta$,
in which the asymmetric structures are more prominent (see Figure \ref{fig:surface_slope}).
This supports our statement in Section \ref{subsec:ep} that
the amount of discrepancy depends on the degree of the asymmetric structures.

The initial slope of surface density also affects the exact resonance location.
The shift in the semi-major axis is larger in the run with $\beta = -1.5$,
a gaseous disk less perturbed by the planet.
This suggests the asymmetric structures in the disk may reduce the exact resonance location.

\begin{figure*}[tbh!]
\plotone{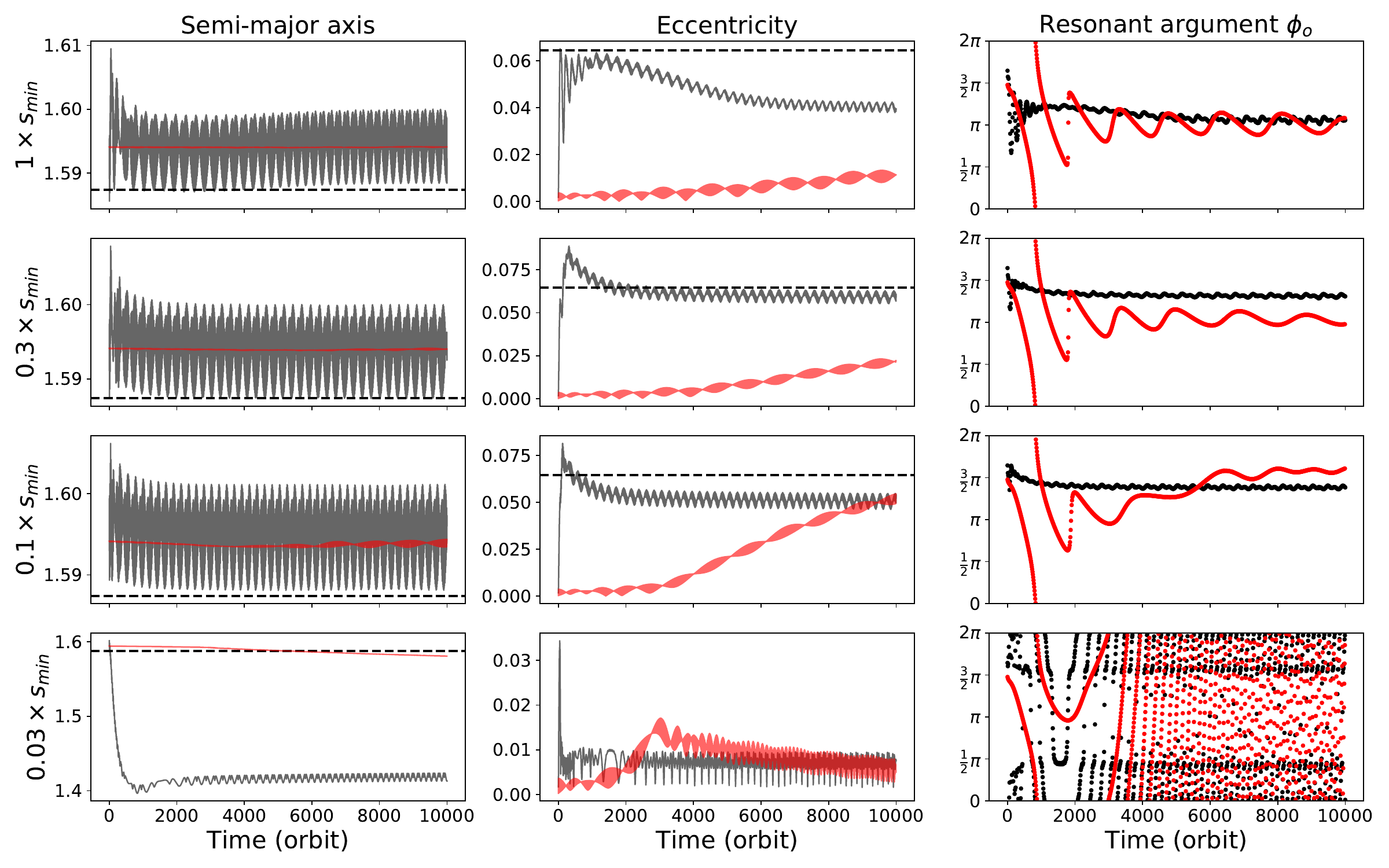}
\caption{Similar to Figure \ref{fig:earth_case},
but with planetesimals of difference sizes placed in near the 1:2 exterior resonance initially.
The results shown in red are from runs with a Earth-sized planet, and in black for runs with a Jupiter-sized planet.
From top to bottom: the orbital elements of planetesimal in sizes of $1$, $0.3$, $0.1$, and $0.03$ times that of $s_{\text{min}}$.
The values of $s_{\text{min}}$ are about $90$ km and $900$ m in the Earth and Jupiter cases, respectively.
The dashed line in the panel of the semi-major axis is the nominal location of 1:2 resonance.
The minimum size of planetesimal that can trigger resonance trapping is about $0.1$ times that of $s_{\text{min}}$.
Planetesimals smaller than this size will pass the 1:2 resonance, and eventually will be halted at the outer edge of the gap.
\label{fig:jupiter_smin}}
\end{figure*}

\subsection{The Minimum Size} \label{subsec:s_min}

The minimum size of planetesimal that can trigger the resonance trapping, $s_{\text{min}}$, 
is derived by taking $\sin\phi_o = -1$, 
which corresponds to $\phi_o = 3\pi / 2$. 
In our simulation results, 
where the planetesimal's size is about two times that of $s_{\text{min}}$,
the equilibrium value of $\phi_o$ is expected to be $7 \pi / 6$ or $11 \pi / 6$, using Eq.\ (\ref{eq:dadt_all}).
However, we note the resonant arguments of planetesimals in MMRs oscillate around $\pi$ instead 
(see the panels for resonant angle in Figures \ref{fig:earth_case} to \ref{fig:jupiter_case}).
This infers the asymmetric structures in the gaseous disk could affect the minimum size predicted by Eq.\ (\ref{eq:s_min}).
In order to study the exact minimum size, 
we perform simulations with planetesimals in sizes of $1$, $0.3$, $0.1$, and $0.03$ times that of $s_{\text{min}}$,
in which the embedded planet is Earth size or Jupiter size.
The values of $s_{\text{min}}$ are about $90$ km and $900$ m for the Earth and Jupiter cases, respectively.
All the planetesimals are placed near the 1:2 exterior resonance initially.
The other initial conditions are same as the corresponding runs.

The results are shown in Figure \ref{fig:jupiter_smin}.
We find the exact minimum size is about 10 times smaller than the value predicted by Eq.\ (\ref{eq:s_min})
in both the Earth and Jupiter cases.
The resonant argument approaches to $3 \pi/ 2$ as the planetesimal's size decreases,
and reaches to about $3 \pi/ 2$ when the size is about $0.1$ times that of $s_{\text{min}}$.
Planetesimals smaller than this size will pass the 1:2 resonance, 
and eventually will be halted by the pressure peak at the outer edge of the gap.

For the Jupiter case, we note the size of the planetesimal also affects the eccentricity.
The planetesimal's eccentricity can reach to the theoretical values in the case of $0.3$ times that of $s_{\text{min}}$,
but are in lower values in cases of larger and smaller sizes.
For planetesimals with smaller size, 
the resonant argument approaches to $3 \pi / 2$,
and both the pumping in eccentricity from resonant perturbations
and damping due to gas drag become stronger.
However, if the drag force is not strong enough to efficiently damp the eccentricity 
excited by the planet during the opposition,
the planetesimal's eccentricity may remain at a higher value,
as in the case of $0.3$ times that of $s_{\text{min}}$.
When the planetesimal's size becomes more smaller, in the case of $0.1$ times that of $s_{\text{min}}$, 
the drag force becomes more stronger, resulting in a smaller equilibrium eccentricity.

On the semi-major axis, the amount of shift is larger for planetesimals with smaller size.
However, the amount differs slightly for planetesimals of different sizes,
with a value of $6 \times 10^{-4}$ in $\Delta$ in the Jupiter case. 
This supports our guess in Section \ref{subsec:ap} 
that the drag force does not contribute to the shift in resonance location significantly.
However, this could be due to that the planetesimals in our simulations are kilometer-sized or subkilometer-sized objects,
in which their dynamics are less perturbed by the gas drag.
For cases of meter-sized pebbles that are well coupled to the gas phase,
the contribution from gas drag in the shift of resonance location may be important.

\section{Discussion and conclusion} \label{sec:discussion}

We implement the drag law into FARGO3D 
and perform long-term 2D hydrodynamic simulations to study the dynamics 
of planetesimals under the influences of gas drag and resonant perturbations.
In this paper, we focus on the $j:(j + 1)$ first-order, exterior MMRs.
We find the exact resonance location is larger than the nominal resonance location.
The shift in resonance location is caused by the precession of the planetesimal,
which is induced mainly by the resonant perturbations from the inner planet, 
instead of the gas drag or the gravitational forces from an axisymmetric gaseous disk.
The amount of shift increases in a gaseous disk less perturbed by the planet.
This suggests the asymmetric structures induced by the planet could affect the exact resonance location,
possibly via the wake--planet interaction.

We reexamine the analytical expressions of the equilibrium eccentricity
and the minimum size of planetesimal that can trigger resonance trapping, outlined in \citet{1985Icar...62...16W},
with our results from hydrodynamic simulations.
We find the expression in \citet{1985Icar...62...16W} underestimates the equilibrium eccentricity
for cases of $j < 5$.
The deviation in eccentricity is significant in particular in the case of 1:2 resonance, 
with a factor of $30 - 40\%$.
The correct eccentricity can be approximated by the modified expression in Eq.\ (\ref{eq:e_eq_modified}),
with an error up to $3\%$ for $j \le 5$.

With the equilibrium eccentricity predicted by our modified formula, 
we find the planetesimal's eccentricity is reduced significantly 
in a gaseous disk in which the asymmetric structures are prominent.
The discrepancy in eccentricity depends on the planet's mass, 
the initial slope of the surface density, and $j$.
For cases of Earth-sized planets or protoplanets, 
where the disk is less disturbed, the discrepancy is negligible, 
and the planetesimal's eccentricity can reach to the theoretical values predicted by the modified formula.
The higher equilibrium eccentricity of planetesimals in the 1:2 resonance 
could promote the destructive collisions between resonant and nonresonant planetesimals.
The resulting fragments could be small enough to pass the resonance trapping
and be accreted onto the planet \citep{1985Icar...62...16W},
or cross the planet into the interior orbits rather than being accreted \citep{1993Icar..106..288K}.
However, the time required for the planetesimal's eccentricity to reach the equilibrium state 
is quite long for the 1:2 resonance, 
indicating the collision rate could be low for planetesimals just being captured into resonances.

For cases of gaseous planets, 
the influences from the asymmetric gaseous profile in the eccentricity become prominent.
The equilibrium eccentricity can be $0.01-0.02$ smaller in value than the theoretical values.
The lower eccentricity thus reduces the relative velocity between resonant and nonresonant planetesimals,
and could be beneficial to the formation of secondary planet.

On the minimum size of planetesimal that can trigger the resonance trapping,
we found the exact size is about $10$ times smaller than the value predicted by \citet{1985Icar...62...16W}
in both the Earth and Jupiter cases.
The factor $10$ seems to be independent or less dependent on the planetary mass and the degree the gaseous disk has been disturbed.
The smaller minimum size could reduce the amount of planetesimals migrated into the planet's feeding zone,
and decrease its growth rate.

Note the results presented in this study are valid for the parameter space we explored.
In a situation where the drag force is weak (e.g., low disk surface density or large planetesimal size),
the damping timescale of eccentricity can be longer than the viscous time.
As the gaseous disk will disperse and the planetesimal may grow via collisions with time, 
the drag force will be further reduced,
and hence the planetesimal's eccentricity may not reach the equilibrium state and keep oscillating 
after the dispersal of the gaseous disk.

\acknowledgments 
We thank the anonymous referee for detailed comments.
This work is supported by the Ministry of Science and Technology, Taiwan, 
through grants MOST 105-2119-M-007-029-MY3 and 106-2112-M-007-006-MY3.

\software{IPython \citep{https://doi.org/10.1109/MCSE.2007.53}, 
		  Matplotlib \citep{https://doi.org/10.1109/MCSE.2007.55}, 
		  NumPy \citep{https://doi.org/10.1109/MCSE.2011.37}, 
		  SciPy \citep{jones2001scipy}, 
		  SymPy \citep{https://doi.org/10.7717/peerj-cs.103}}

\appendix

\section{Evolution of Surface Density over Viscous Time} \label{appendix_a}

For most simulations presented in this study, 
the simulation time is comparable to or longer than the viscous timescale,
which is $r^2 / \nu \sim 1.6 \times 10^4$ orbits at $r = 1$.
The viscous evolution of the gaseous disk will redistribute the density profile,
and may affect the theoretical value of equilibrium eccentricity due to the change in the density slope.

\begin{figure*}[!tb]
\plotone{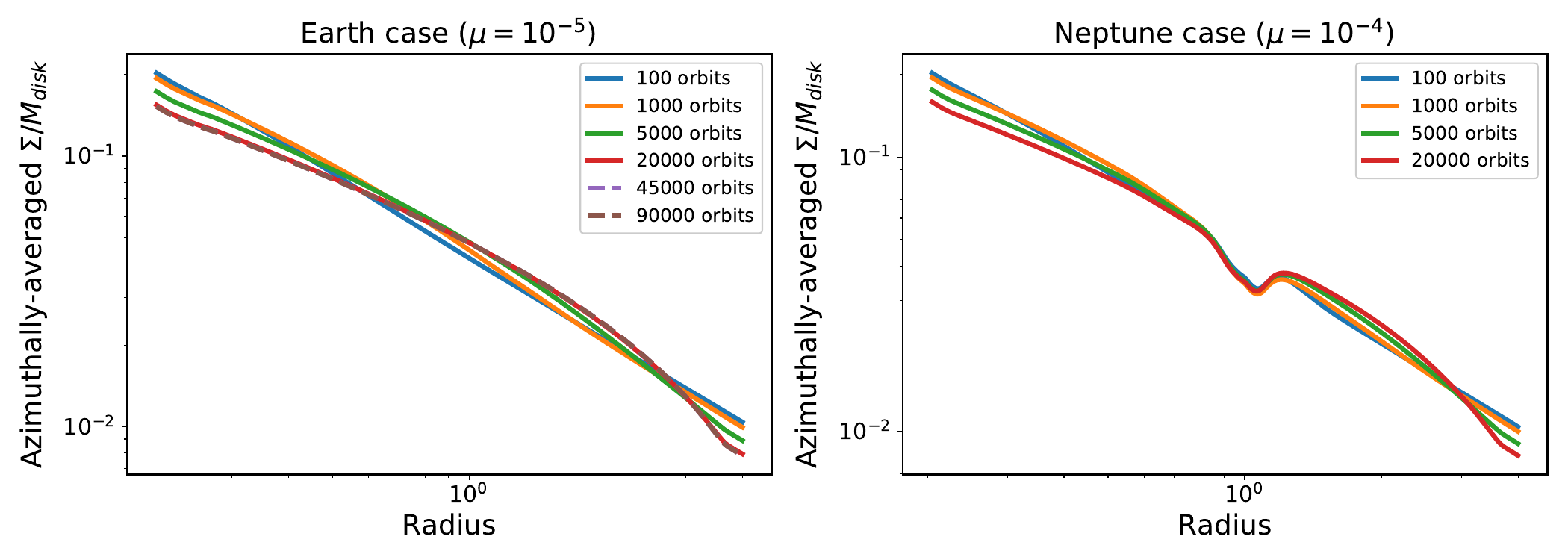}
\caption{Azimuthally averaged surface density at different simulation times.
The left panel shows the evolution of surface density of the simulation with an Earth-sized planet,
and the right panel shows the simulation with a Neptune-sized planet.
\label{fig:dens_evo}}
\end{figure*}

In Figure \ref{fig:dens_evo}, we plot the azimuthally averaged surface density at different simulation times
for the Earth and Neptune cases presented in Figures \ref{fig:earth_case} and \ref{fig:neptune_case}.
Within one viscous time, the density in the inner disk is smoothed out and becomes flatter.
For the outer disk, the gases are accumulated outside the planet ($r = 1$), 
and the density profile near the nominal resonance locations also becomes flatter.

After one viscous time, the disk profile stops evolving.
We think this is caused by the damping boundary condition used in this study.
The damping boundary condition is employed to reduce the reflecting wave near the boundary 
by recovering the cells near the boundary to the initial state. 
In other words, it also serves as if there are additional mass sources near the outer boundary, 
which replenish the mass loss due to accretion processes 
and could halt the further viscous evolution due to mass loss.

However, the change in the density slope does not significantly reduce 
the theoretical value of equilibrium eccentricity (see Figure \ref{fig:numerical_Adachi}).
In the Neptune case, we find the small equilibrium eccentricity can not be explained 
by the theoretical values even using the final density slope.

\listofchanges

\end{document}